\documentclass[
reprint,
superscriptaddress,
frontmatterverbose, 
showpacs,
preprintnumbers,
nofootinbib,
amsmath,
amssymb,
aps,
floatfix,
twocolumn
]{revtex4-2}

\usepackage[inline]{enumitem}
\usepackage[utf8]{inputenc}
\usepackage[normalem]{ulem}
\usepackage{graphicx}% Include figure files
\usepackage{dcolumn}% Align table columns on decimal point
\usepackage{bm}% bold math
\usepackage{color}
\usepackage[dvipsnames]{xcolor}
\usepackage[
    colorlinks = true,
    linkcolor = BlueViolet,
    anchorcolor = purple,
    citecolor = purple,
    filecolor = purple,
    urlcolor = BlueViolet]{hyperref}
    
\usepackage{url}
\usepackage{xspace}
\usepackage{slashed}
\usepackage{multirow,bigstrut}
\usepackage{mathrsfs} % pretty maths
\usepackage{relsize,amsmath}
\usepackage[geometry]{ifsym}
\usepackage{amssymb}% http://ctan.org/pkg/amssymb
\usepackage{pifont}% http://ctan.org/pkg/pifont
\usepackage{physics}

\usepackage{cleveref}

\usepackage{rotating}
\usepackage{ragged2e}

\newcommand{\apjl}{Astrophys. J. Lett.\,}

\usepackage{fontawesome} % for code link icons
\definecolor{blue-violet}{rgb}{0.33, 0.17, 0.89}

\def\miniboone{MiniBooNE\xspace}

\renewcommand{\phi}{\varphi}

\newcounter{CommentCount}
\definecolor{MH}{rgb}{0.0,0.6,9}
\definecolor{palatinate}{rgb}{0.494, 0.192, 0.482}

\definecolor{teal}{HTML}{008080}

\usepackage{siunitx}
\DeclareSIUnit \s {\second}
\DeclareSIUnit \ns {\nano\second}
\DeclareSIUnit \mus {\micro\second}
\DeclareSIUnit \ms {\milli\second}
\DeclareSIUnit \MB {\mega\byte}
\DeclareSIUnit \GB {\giga\byte}
\DeclareSIUnit \TB {\tera\byte}
\DeclareSIUnit \PB {\peta\byte}
\DeclareSIUnit \Mbps {\mega\bit/\s}
\DeclareSIUnit \Gbps {\giga\bit/\s}
\DeclareSIUnit \Tbps {\tera\bit/\s}
\DeclareSIUnit \Pbps {\peta\bit/\s}
\DeclareSIUnit \kton {\kilo\tonne} % changed  back to kton
\DeclareSIUnit \kt {\kilo\tonne}
\DeclareSIUnit \Mt {\mega\tonne}
\DeclareSIUnit \eV {\electronvolt}
\DeclareSIUnit \keV {\kilo\electronvolt}
\DeclareSIUnit \MeV {\mega\electronvolt}
\DeclareSIUnit \GeV {\giga\electronvolt}
\DeclareSIUnit \TeV {\tera\electronvolt}
\DeclareSIUnit \PeV {\peta\electronvolt}
\DeclareSIUnit \EeV {\exa\electronvolt}
\DeclareSIUnit \m {\meter}
\DeclareSIUnit \cm {\centi\meter}
\DeclareSIUnit \in {\inchcommand}
\DeclareSIUnit \km {\kilo\meter}
\DeclareSIUnit \kV {\kilo\volt}
\DeclareSIUnit \kW {\kilo\watt}
\DeclareSIUnit \MW {\mega\watt}
\DeclareSIUnit \MHz {\mega\hertz}
\DeclareSIUnit \mrad {\milli\radian}
\DeclareSIUnit \year {years}
\DeclareSIUnit \POT {POT}
\DeclareSIUnit \sig {$\sigma$}
\DeclareSIUnit\parsec{pc}
\DeclareSIUnit\lightyear{ly}
\DeclareSIUnit\foot{ft}
\DeclareSIUnit\ft{ft}
\DeclareSIUnit \ppb{ppb}
\DeclareSIUnit \ppt{ppt}
\DeclareSIUnit \samples{S}
\DeclareSIUnit \pe{PE}
\DeclareSIUnit \T{T}

\newcommand{\enu}{\E_\enu}

\begin{document}

%\preprint{\hfill FTPI-MINN-??-??}
%\preprint{some fermi number}
\title{Exploring the dark sectors via the cooling of  white dwarfs}

\author{Jaime Hoefken Zink}
\email{jaime.hoefkenzink2@unibo.it}
\affiliation{Dipartimento di Fisica e Astronomia, Universit\`a di Bologna, via Irnerio 46, 40126 Bologna, Italy}
\affiliation{INFN, Sezione di Bologna, viale Berti Pichat 6/2, 40127 Bologna, Italy}

\author{Maura E. Ramirez-Quezada}
\email{me.quezada@hep-th.phys.s.u-tokyo.ac.jp}
\affiliation{Department of Physics, University of Tokyo, Bunkyo-ku, Tokyo 113--0033, Japan,}
\affiliation{ {Dual CP Institute of High Energy Physics, C.P. 28045, Colima, M\'exico}}

\date{\today}

\begin{abstract}
As dense and hot bodies with a well-understood equation of state, white dwarfs offer a unique opportunity to investigate new physics. In this  paper, we examine the role of dark sectors, which are extensions of the Standard Model of particle physics that are not directly observable, in the cooling process of white dwarfs. Specifically, we examine the role of a dark photon, within the framework of a three-portal Model, in enhancing the neutrino emission during the cooling process of white dwarfs.  We compare this scenario to the energy release predicted by the Standard Model. By analyzing the parameter space of dark sectors,  our study aims to identify regions that could lead to significant deviations from the expected energy release of white dwarfs.
\end{abstract}

\maketitle
%%%%%%%%%%%%%%%%%%%%%%%%%%%%%
\section{Introduction}
%%%%%%%%%%%%%%%%%%%%%%%%%%%%% 
Dark sectors (DS) have gained increasing importance in recent years as a potential extension to the limitations of the Standard Model of particle physics.
 In particular, several theoretical extensions have been proposed to explain the origin of neutrino masses and mixing, with the Type-I seesaw mechanism and its variations receiving significant attention.  These models only introduce  heavy-neutral leptons (HNLs) as part of their minimal framework~\cite{Asaka:2005pn}. However, a more compelling setup has recently been proposed as a potential solution, where the HNLs are considered as part of a richer low-energy dark sector~\cite{Pospelov:2011ha, Harnik:2012ni,Batell:2016zod,Farzan:2016wym,DeRomeri:2017oxa,Magill:2018jla,Bertuzzo:2018ftf,Bertuzzo:2018itn,Ballett:2019cqp,Ballett:2019pyw,Coloma:2019qqj}.
The non-minimal  DS introduces a new particle called the ``dark photon'', which can interact with both electromagnetic and dark currents. One interesting feature of this kind of model  is that  it could also  explain many of the experimental anomalies which cannot be accounted for just by appealing to the SM, such as \miniboone low energy excess~\cite{Ballett:2018ynz,Bertuzzo:2018itn,Abdullahi:2020nyr,Datta:2020auq,Dutta:2020scq,Abdallah:2020biq,Abdallah:2020vgg} and the muon $(g-2)$ anomaly~\cite{Abdullahi:2020nyr,Dutta:2020scq,Abdallah:2020biq,Abdallah:2020vgg,Abdullahi:2023tyk}, among other ones.   To fully explore the potential and test the predictions of the three-portal model, further theoretical investigations are necessary. However, due to the introduction of several new parameters and particles, experimental testing of the model may become more complex and challenging.

White dwarfs (WD) provide unique opportunities to study new physics due to their extreme conditions.  By analyzing the luminosity emitted by white dwarfs, it is feasible to probe the underlying physics in ways that are impossible in Earth-based laboratories. One particular aspect is the cooling process of young and hot white dwarfs, which is primarily driven by plasmon decay into neutrinos originating from their core~\cite{Kantor:2007kf,Winget:2004}. 
The potential for enhanced cooling through plasmon decay, mediated by new particles, has been explored in the context of models with anomaly-free symmetry, such as U(1)$_{B-L}$ and U(1)$_{L_\alpha-L_\beta}$. These decay processes contribute to the production of the SM  neutrinos~\cite{Bauer:2018onh} and could also lead to the generation of new light particles~\cite{Dreiner:2013tja}. Furthermore
some studies have shown that there could be other important mechanisms under the assumption of a strong magnetic field~\cite{1970Ap&SS...7..407C,1970Ap&SS...9..453C,DeRaad:1976kd,skobelev1976reaction,galtsov1972photoneutrino,Kennett:1999jh}, such as neutrino pair synchrotron emission from electrons~\cite{Landstreet:1967zz,chaudhuri1970neutrino,1981AN....302..167I,Kaminker:1992su,Bhattacharyya:2005tf,Drewes:2021fjx}.

The aim of this study is to investigate the emission of neutrinos from white dwarfs, with a focus on beyond the SM (BSM) interactions in the neutrino production rate. Specifically, we will use the  DS model, known as the ``three-portal'' model,  proposed in Ref.~\cite{Abdullahi:2020nyr}. Since the dark photon is capable of interacting with neutrinos as well as  electromagnetic-charged particles, its presence could potentially alter the production of neutrinos through plasmon decay at the early stages of a WD. 

The structure of this paper is organized as follows. In Section~\ref{sec:frammework}, we provide a concise overview of the physics of white dwarfs. Additionally, we introduce the three-portal model that we are considering in this study, along with its key features. In Section~\ref{sec:emissionrates}, we present the emission rate of neutrino production, both within the context of SM interactions and with the modified expressions that include the DS. Our results are discussed and presented in Section~\ref{sec:results}, followed by our conclusions in Section~\ref{sec:conclusions}.

%%%%%%%%%%%%%%%%%%%%%%%%%%%%%
\section{Theoretical framework}
\label{sec:frammework}
%%%%%%%%%%%%%%%%%%%%%%%%%%%%%

\subsection{White dwarfs}
A white dwarf is a dense star that forms after a normal star has exhausted its nuclear fuel and undergone the final stages of its evolution. This process ejects the outer layers of the star, leaving behind a hot, dense core primarily composed of carbon and oxygen. The core is supported by electron degeneracy pressure, which prevents it from collapsing further and instead causes it to contract and cool over billions of years. White dwarfs are incredibly dense ($\sim10^6\,\rm kg/m^3$), with a mass similar to the Sun but a size similar to that of the Earth.

Electrons play a crucial role in determining the equation of state (EoS) and the structure of white dwarfs.  In~Ref.~\cite{salpeter61}, a theoretical EoS for WDs is obtained by assuming a Wigner-Seitz (WS) cell, which is a uniformly negatively charged spherical cell with a positively charged ion at its centre. Corrections to the electron energy due to the electrostatic potential were also introduced, which are the most significant correction to the EoS. Further corrections, such as considering the non-rigidity of ions within the WS cell and self-interactions, were also taken into account. However, it was determined that these corrections do not significantly affect the behaviour of high-density matter. 

Initially, white dwarfs have extremely high temperatures before cooling down to become faint objects. There are several stages of cooling that it goes through, with each stage characterized by a different mechanism of energy loss.  The equation governing the temperature evolution of WDs depends on the cooling mechanism and the physical properties,
\begin{equation}
    \frac{dT_\star}{dt}=-\frac{L_\gamma}{4 \pi R_\star \sigma_{\rm SB}T_\star}-\frac{L_\nu}{4 \pi R_\star \sigma_{\rm SB}T_\star}.
\end{equation}
The left-hand side represents the temperature change rate with respect to time, where $T_\star$ is the temperature of the white dwarf. The right-hand side consists of two terms, representing the rates of energy loss due to photon radiation ($L_\gamma$) and neutrino emission ($L_\nu$), respectively. Here $R_\star$ is the WD radius, and $\sigma_{\rm SB}$ is the Stefan-Boltzmann constant.

The dominant cooling mechanism for newborn WDs is through the emission of neutrinos, produced primarily through plasmon decay~\cite{Kantor:2007kf,Winget:2004}. The produced  neutrinos  easily escape the dense core of the WD, carrying away energy and facilitating the loss of thermal energy.
As the white dwarf continues to cool down, its temperature eventually drops to approximately $10^3 \ \rm K$, at which point it enters the photon cooling stage. In this stage, the white dwarf radiates energy primarily in the form of photons, and its luminosity is dominated by photon radiation. Consequently, this stage is not of interest in the discussion that follows.

%%%%%%%%%%%%%%%%%%%%%%%%%%%%%
\subsection{Three portal model}
%%%%%%%%%%%%%%%%%%%%%%%%%%%%%

In this section, we briefly describe the comprehensive three-portal model that extends the SM of particle  physics by introducing a new $U(1)$ symmetry known as $U(1)_X$, which spontaneously breaks at the sub-GeV scale. The new symmetry is accompanied by a Higgs singlet ($\Phi$), a gauge field mediator ($X^\mu$), sterile neutrinos ($N$), and dark neutrinos ($\nu_D$).  The 
interactions between the new particles and the SM particles are described by the Lagrangian~\cite{Abdullahi:2020nyr},

\begin{align}
\mathscr{L} &= \mathscr{L}_{\mathrm{SM}} + (D^x_\mu \Phi)^\dagger (D^{x \mu} \Phi) - V(\Phi,H) - \frac{1}{4} X_{\mu \nu} X^{\mu \nu} \nonumber\\ 
&- \frac{\sin \chi}{2} B_{\mu \nu} X^{\mu \nu} + \overline{N} i \slashed{\partial}N + \overline{\nu}_D i \slashed{D}^x \nu_D\nonumber \\ 
&- [y_\nu^\alpha(\overline{L}_\alpha\! \cdot \!\widetilde{H})N^C\! + \frac{\mu'}{2} \overline{N}N^C\! + y_N \overline{N} \nu_D^C \Phi + \mathrm{h.c.}],
\end{align}
where $\widetilde{H} \equiv i \sigma_2 H^*$, $D^x_\mu \equiv \partial_\mu - i g_D X_\mu$, $\sin \chi$ is a small coupling for the kinetic mixing between the hypercharge and the gauge field $X^\mu$. The terms in square brackets represent the neutrino mass terms and interactions with the Higgs. 

The three-portal model incorporates three possible communication channels between the SM and the DS. 
These portals allow for interactions and exchanges of particles between the two sectors. The first portal is the scalar portal, which occurs through the mixing of the DS Higgs singlet with the SM Higgs doublet. This mixing creates a scalar field that can interact with SM and DS particles. The second portal is the neutrino portal, which occurs through the mixing of SM neutrinos with dark neutrino states. Finally, we have a vector portal, which appears thanks to the kinetic mixing of the $X$ field with the hypercharge, $B$, and through a broken symmetry resulting in a dark photon $Z^\prime$. This new particle can interact with both electromagnetic and dark currents. 

In the three-portal model, the  mass of the dark photon is typically assumed to be less than $\mathcal{O}(1 \mathrm{ GeV})$. When the mass of the dark photon is much smaller than the mass of the $Z$ boson, $M_{Z^\prime}/M_Z\ll1$, the complicated interactions of the dark photon can be approximated by a simplified form as follows,
\begin{equation}
\label{darkphotoninteraction}
\begin{split}
\mathcal{L}_I \simeq - \epsilon e J_\mu^{\mathrm{EM}} Z^{\prime \mu} - g_{\mathrm{D}} J_\mu^{\mathrm{D}} Z^{\prime \mu},
\end{split}
\end{equation}
where $J_\mu^{\mathrm{EM}}$ is an electromagnetic current, $J_\mu^{\mathrm{D}}$ is a dark current that consists of dark neutrino states: $\bar{\nu}_\mathrm{D} \gamma_\mu \nu_\mathrm{D}$, and $Z^{\prime \mu}$ is a dark photon. We consider $\epsilon$ a small number, while $g_{\mathrm{D}}$ is not highly constrained. 

Once the electroweak and dark symmetries are spontaneously broken, and considering $\nu_\alpha$, $N$ and $\nu_D$ to be the SM-flavor, sterile and dark states, respectively, the neutrino mass matrix takes a form similar to an inverse~\cite{Mohapatra:1986bd,Gonzalez-Garcia:1988okv} or an extended seesaw~\cite{Barry:2011wb, Zhang:2011vh}:  
\begin{equation}
\begin{split}
\mathscr{L}_{\mathrm{mass}}^\nu  = -\frac{1}{2}\begin{pmatrix} \overline{\nu_\alpha} &\overline{N} &\overline{\nu_D} \end{pmatrix} \begin{pmatrix}
0_{3 \times 3} &m_D^\mathrm{T} &0 \\m_D &\mu' &\Lambda^\mathrm{T} \\0 &\Lambda &0
\end{pmatrix}
\begin{pmatrix} \nu_\alpha \\N^C \\\nu_D^C \end{pmatrix}
\end{split}
\end{equation}
where we define,
\begin{equation}
m_D = \Bigg[\frac{y_\nu^\alpha}{\sqrt{2}} v_H\Bigg]^\mathrm{T},\,\, \ \Lambda = \Bigg[\frac{y_N}{\sqrt{2}} v_\phi\Bigg]^\mathrm{T}
\end{equation}
$v_{H}$ and $v_\phi$ correspond to the SM Higgs vacuum expectation value (VEV) and the dark scalar  VEV, respectively. Each vector runs over $\alpha$ (neutrino SM-flavor states) and $N$ (neutrino sterile states). The mass matrix can be diagonalized using \footnote{In order to account for the cancellation of the chiral anomaly, the number of dark neutrino states should consider the pairing of right-handed and left-handed dark states. One way to achieve this is by introducing Dirac terms in the mass matrix, which would preserve the chiral symmetry. If Majorana terms were included instead, they would explicitly break the dark gauge symmetry.
It is important to note that our calculations remain independent of the precise approach taken in this regard}
\begin{equation}
\begin{split}
\widehat{M} &= \begin{pmatrix} U_\alpha^\mathrm{T} &U_N^\mathrm{T} &U_D^\mathrm{T} \end{pmatrix} \begin{pmatrix}
0_{3 \times 3} &m_D^\mathrm{T} &0 \\m_D &\mu' &\Lambda^\mathrm{T} \\0 &\Lambda &0
\end{pmatrix} \begin{pmatrix} U_\alpha \\U_N \\U_D \end{pmatrix} \\&= U^\mathrm{T} M U.
\end{split}
\end{equation}
The strength of the  dark photon interaction with active neutrinos  is proportional to the components of the  mixing matrix $U$ as $\propto g_D U_{Di}^\ast U_{Dj}\gamma_\nu P_{\rm L}$. Notice that the standard PMNS neutrino mixing matrix is typically obtained by examining the flavour and light neutrino sectors. 
%%%%%%%%%%%%%%%%%%%%%%%%%%%%%
\section{Neutrino emission rate}
\label{sec:emissionrates}
%%%%%%%%%%%%%%%%%%%%%%%%%%%%%

%%%%%%%%%%%%%%%%%%%%%%%%%%%%%
\subsection{Review of emission rate}
%%%%%%%%%%%%%%%%%%%%%%%%%%%%%

The release of neutrinos  has a significant impact on the energy loss of stars that are extremely hot or dense. The production rate of neutrinos can be greatly affected by the combined effects of the stellar plasma. For instance, photons may decay into pairs of neutrinos ($\gamma\to\nu\overline{\nu}$), carrying away energy~\cite{Kantor:2007kf}. This is made possible by modifying the dispersion relations of the photon due to thermal effects that allow it to decay. 

Throughout this section and in subsequent discussions, we will follow the convention of calling ``photon" the transverse polarisation, while ``plasmon" is the longitudinal one~\cite{Braaten:1993jw}.

\subsubsection{Photon-self energy at finite temperature}
To determine the neutrino emissivity due to plasmon decay, we must first compute the self-energy of the photon at a finite temperature, as it plays a crucial role in the emission rate computation. The most comprehensive expression for the photon-self energy at a finite temperature is given as~\cite{Braaten:1993jw,lebellac_1996},  
\small{
\begin{equation}
 \begin{split}
\Pi^{\mu \nu}& = 4e^2 \int \frac{d^3 k}{(2\pi)^3} \frac{f_e(E_k) + f_{\overline{e}}(E_k)}{2E_k} \\&\times \frac{Q \cdot K(K^\mu Q^\nu + K^\nu Q^\mu) - Q^2 K^\mu K^\nu - (Q \cdot K)^2 g^{\mu \nu}}{(Q \cdot K)^2 - Q^4/4}
\end{split}
\end{equation}}
here, $Q = (q_0, \vec{q})$ and $K=(E_k,\vec{k})$ represent the 4-momentum of the photon and electron (or positron), respectively and  $Q\cdot K= q_0 E_k - \vec{k}\cdot \vec{q}$. The thermal distribution of the electron (or positron) is denoted by $f_e(E_k)$ (or $f_{\overline{e}}(E_k)$). The self-energy tensor of a thermal photon can be divided into two components: longitudinal and transverse components as follows,
\begin{equation}
\Pi^{\mu \nu} = F P_L^{\mu \nu} + G P_T^{\mu \nu}\label{eq:photon_self_energyTL}
\end{equation}
where the projectors are:
\begin{equation}
\begin{split}
P^{\mu \nu}_T &= \big( \delta^{ij} - \hat{q}^i \hat{q}^j\big) \delta^\mu_i \delta^\nu_j \\
P^{\mu \nu}_L &= \bigg( -g^{\mu \nu} + \frac{Q^\mu Q^\nu}{Q^2} \bigg) - P^{\mu \nu}_T .
\end{split}
\end{equation}
Evaluating the $00$-elements of the longitudinal and transverse components of the photon self-energy yields $P_T^{00}=0$ and $P_L^{00}=\vec{q}^{\,2}/Q^2$, respectively. Consequently, we find that $F=Q^2/\vec{q}^{\,2}\Pi^{00}$. Similarly, considering the $xx$-elements of the transverse and longitudinal components gives $P_T^{xx}=1$ and $P_L^{xx}=0$, respectively, leading to $G=\Pi^{xx}$. As a result, we can express Eq.~\eqref{eq:photon_self_energyTL} as follows,
\begin{equation}
    \Pi^{\mu\nu}=\frac{Q^2}{\vec{q}^{\,2}}\Pi^{00} P_L^{\mu\nu}+ \Pi^{xx}P_T^{\mu\nu}.\label{eq:photon_self-energy}
\end{equation}
We identify the first and second  terms as $\Pi_L^{\mu\nu}$ and $\Pi_T^{\mu\nu}$, respectively.

It is necessary to determine various factors to calculate the photon polarization $4$-vectors at finite temperature. This includes the dispersion functions of photons and plasmons as well as the corresponding residual functions ($Z_t(q)$ and $Z_l(q)$), where $q\equiv |\vec{q}|$. In this context, the dispersion relations for photons and plasmons provide the relationship between the frequency $\omega_t$ and $\omega_l$ and the momentum $q$ of the photon/plasmon. These relations  provide information about the energy of the photon/plasmon.

The dispersion function for the plasmon is determined using the longitudinal propagator.  This propagator is found by considering $D^{00} = \frac{1}{q^2 - \Pi_{L}(Q)}$. If $\omega_l (q)$ represents the energy of the longitudinally polarized plasmon on-shell, in its vicinity it can be deduced that~\cite{Braaten:1993jw},
\begin{equation}
\label{long_prop}
\lim_{q_0 \to \omega_l (q)} D^{00} = \frac{\omega_l(q)^2}{q^2} \frac{Z_l(q)}{q_0^2 - \omega_l(q)^2}
\end{equation}
and the dispersion relation for $\omega_l$ is obtained by setting the denominator equal to zero, meaning that $\Pi_{L}(\omega_l(q),q) = q^2$. Therefore, we obtain
\begin{equation}
\omega_l(q)^2 = \frac{\omega_l(q)^2}{q^2} \Pi_L(\omega_l(q),q). 
\end{equation}
Furthermore, since $\omega_l(q)$ is also the pole of the propagator, the value of $Z_l(q)$ can be easily found from Eq.~\eqref{long_prop} 
\begin{equation}
Z_l (q) =  \frac{q^2}{\omega_l(q)^2} \bigg[ - \frac{\partial \Pi_{L}}{\partial q_0^2} (\omega_l(q), q) \bigg]^{-1}\label{eq:Zlong}
\end{equation}

Similarly, for the transverse propagator, where $x$ is a transverse direction, we have $D^{xx} = \frac{1}{q_0^2 - q^2 - \Pi_{T}(Q)}$. The pole of the propagator is given by $\omega_t(q)$, and in its vicinity, the propagator takes the form~\cite{Braaten:1993jw}
\begin{equation}
\lim_{q_0 \to \omega_t (q)} D^{xx} = \frac{Z_t(q)}{q_0^2 - \omega_t(q)^2},\label{eq:propt}
\end{equation}
the dispersion relation for $\omega_t$ is found to be
\begin{equation}
\omega_t(q)^2 = q^2 + \Pi_T(\omega_t(q),q),
\end{equation}
and the residual function  $Z_t(q)$ is given by
\begin{equation}
Z_t (q) = \bigg[ 1 - \frac{\partial \Pi_{T}}{\partial q_0^2} (\omega_t(q), q) \bigg]^{-1}. \label{eq:Ztrans}
\end{equation}

Finally, we can define the photon/plasmon polarization $4$-vectors modified due to the effects of the temperature, 
\begin{equation}
\begin{split}
&\varepsilon^\mu (q, \lambda = 0) = \frac{\omega_l(q)}{q} \sqrt{Z_l(q)} (1,0)^\mu \\
&\varepsilon^\mu (q, \lambda = \pm 1) = \sqrt{Z_t(q)} (0,\varepsilon_\pm (q))^\mu
\end{split}
\end{equation}
where $\varepsilon_\pm (q)$ are two unit mutually orthogonal vectors on the plane transverse to $\vec{q}$.

During the integration of the self-energy, the quantity $v \equiv k/E_k$, where $k\equiv|\vec{k}|$, may lead to  three different  temperature regimes: a non-relativistic, relativistic or degenerate regime. 
In the non-relativistic limit ($T \ll m_e$) where particles are non-degenerate ($T \ll m_e - \mu$), $v$ is equal to $0$. 
On the other hand, in the relativistic limit, where electrons can be considered massless, the parameter $v$ takes on the value of 1. This limit is observed in either a high-density regime ($\mu \gg m_e$) or a high-temperature regime ($T \gg m_e$).
Finally, $v= v_F \equiv p_F / E_F$ in the degenerate limit, where $p_F \equiv \big(3 \pi^2 n_e \big)^{1/3}$ is the Fermi momentum, $E_F \equiv \mu(T=0)$ is the Fermi energy, and $n_e$ is the number density of electrons. 
In the degenerate limit, $df_e/dk$ peaks at $v_F $. In such cases, we can approximate the integrals by extracting this central value, denoted as $v_*$, from the integral so that~\cite{Braaten:1993jw},
\begin{align}
\Pi_{L} &= \omega_p^2  \frac{3}{v^2_*} \Bigg( \frac{q_0}{2 v_* q} \ln \frac{q_0 + v_* q}{q_0 - v_* q} - 1 \Bigg) \\
\Pi_{T} &= \omega_p^2  \frac{3}{2v^2_*} \Bigg( \frac{q_0^2}{q^2} - \frac{q_0^2 - v_*^2 q^2}{q^2} \frac{q_0}{2 v_* q} \ln \frac{q_0 + v_* q}{q_0 - v_* q} \Bigg)\\
\omega_p^2 &= \frac{4\alpha}{\pi} \int_0^{\infty} dk \frac{k^2}{E_k} \bigg(1 - \frac{1}{3} v^2\bigg) \big(f_e(E_k) + f_{\overline{e}}(E_k)\big),
\end{align}here, $\omega_p$ refers to the plasma frequency, which describes the  oscillation of the plasma itself due to changes in the separation of electric charges caused by the motion of the charged particles. This type of oscillation is also referred to as Langmuir waves~\cite{langmuir1928oscillations}.  By using the aforementioned approximations in Eq.~\eqref{eq:Zlong} and \eqref{eq:Ztrans}, we can express $Z_l$ and $Z_t$ as follows,
\begin{align}
Z_{l}^{-1} &= \frac{3 \omega_p^2}{2 v^2_* q^2} \Bigg( \frac{\omega_l^2}{\omega_l^2 - v_*^2 q^2} - \frac{\omega_l}{2 v_* q} \ln \frac{\omega_l + v_* q}{\omega_l - v_* q} - 1 \Bigg) \\
Z_{t}^{-1} &= 1 - \frac{3 \omega_p^2}{2v^2_* q^2} \Bigg( \frac{3}{2} - \frac{3 \omega_t^2 - v_*^2 q^2}{2 \omega_t^2} \frac{\omega_t}{2 v_* q} \ln \frac{\omega_t + v_* q}{\omega_t - v_* q} \Bigg)
\end{align}

In the computation of the photon/plasmon decay, an additional contribution to the self-energy arises from a diagram containing a $\gamma^5$ matrix at one of its vertices. This axial contribution is due to the electroweak sector of neutrinos and must also be calculated. The magnitude of this purely transverse tensor is given by~\cite{Braaten:1993jw}
\begin{align}
\Pi_{A} (Q) = &\frac{2 \alpha}{\pi} \frac{Q^2}{q} \int dk \frac{k^2}{E_k^2} \big[ f_e (E_k) - f_{\overline{e}} (E_k) \big]\nonumber \\[0.1cm]
& ~~~~~~~~~~~\times \bigg(\frac{q_0}{2 q v} \ln \frac{q_0 + v q}{q_0 - v q} - \frac{Q^2}{q_0^2 - v^2 q^2}\bigg)
\end{align}
Here $\alpha$ is the fine structure constant. By employing the same approximation as previously mentioned, we can obtain the following expression
\begin{equation}
\Pi_A (Q) = \omega_A \frac{Q^2}{q} \frac{3}{v_*^2} \bigg( \frac{q_0}{2 q v_*} \ln \frac{q_0 + v_* q}{q_0 - v_* q} - 1 \bigg).
\end{equation}
Here, $\omega_A$ represents an axial frequency, which is defined as,
\begin{equation}
\begin{split}
\omega_A &= \lim_{q \to 0} \frac{\Pi_A (\omega_t(q),q)}{q}
\\&=-\frac{2 \alpha}{3\pi} \int dk \frac{k^3}{E_k^2} \frac{d}{dk} \big[ f_e (E_k) - f_{\overline{e}} (E_k) \big].
\end{split}
\end{equation}

We now have all the necessary elements to compute the plasmon decay and the neutrino emission rates. 
\subsubsection{Amplitude and decay width}
\begin{figure}
    \centering
    \includegraphics[width=0.3\textwidth]{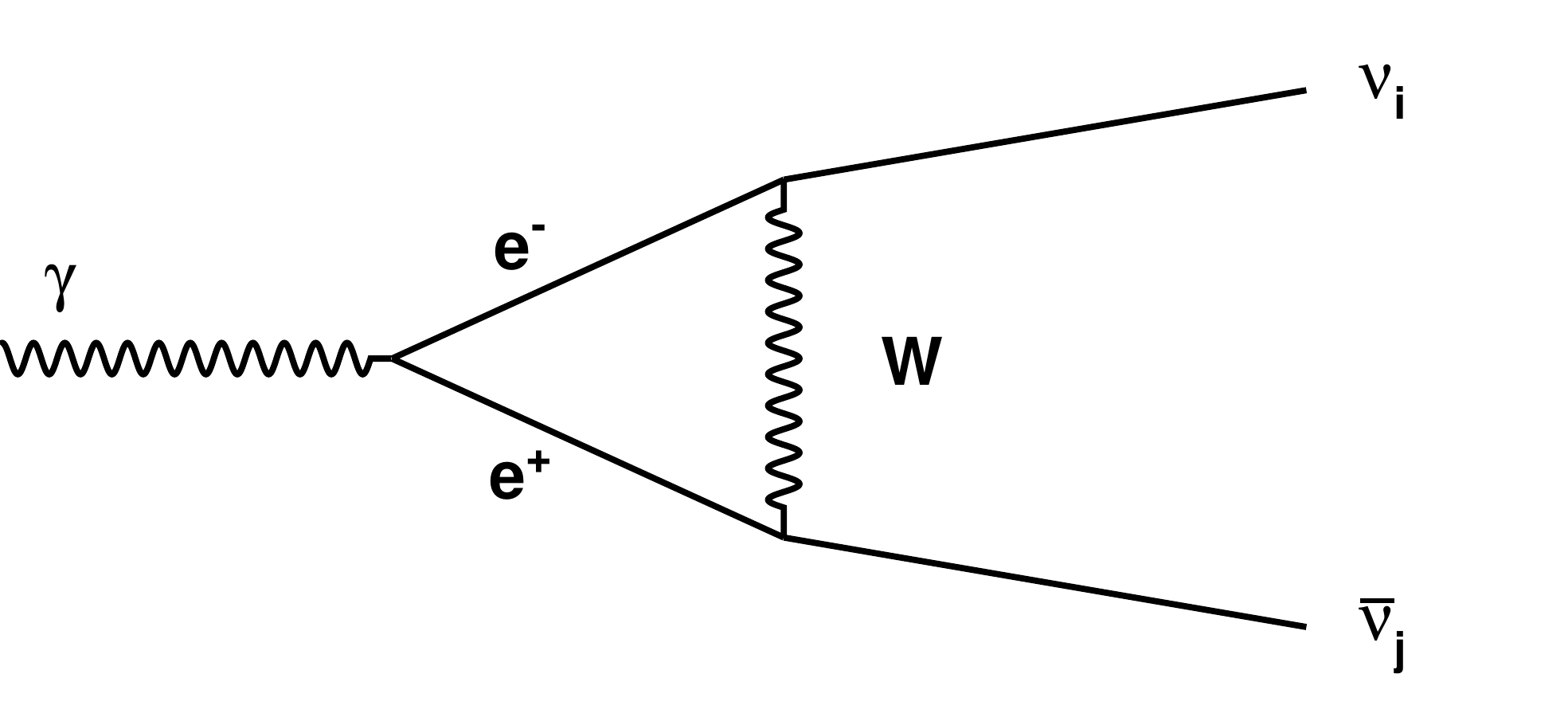}
    \includegraphics[width=0.3\textwidth]{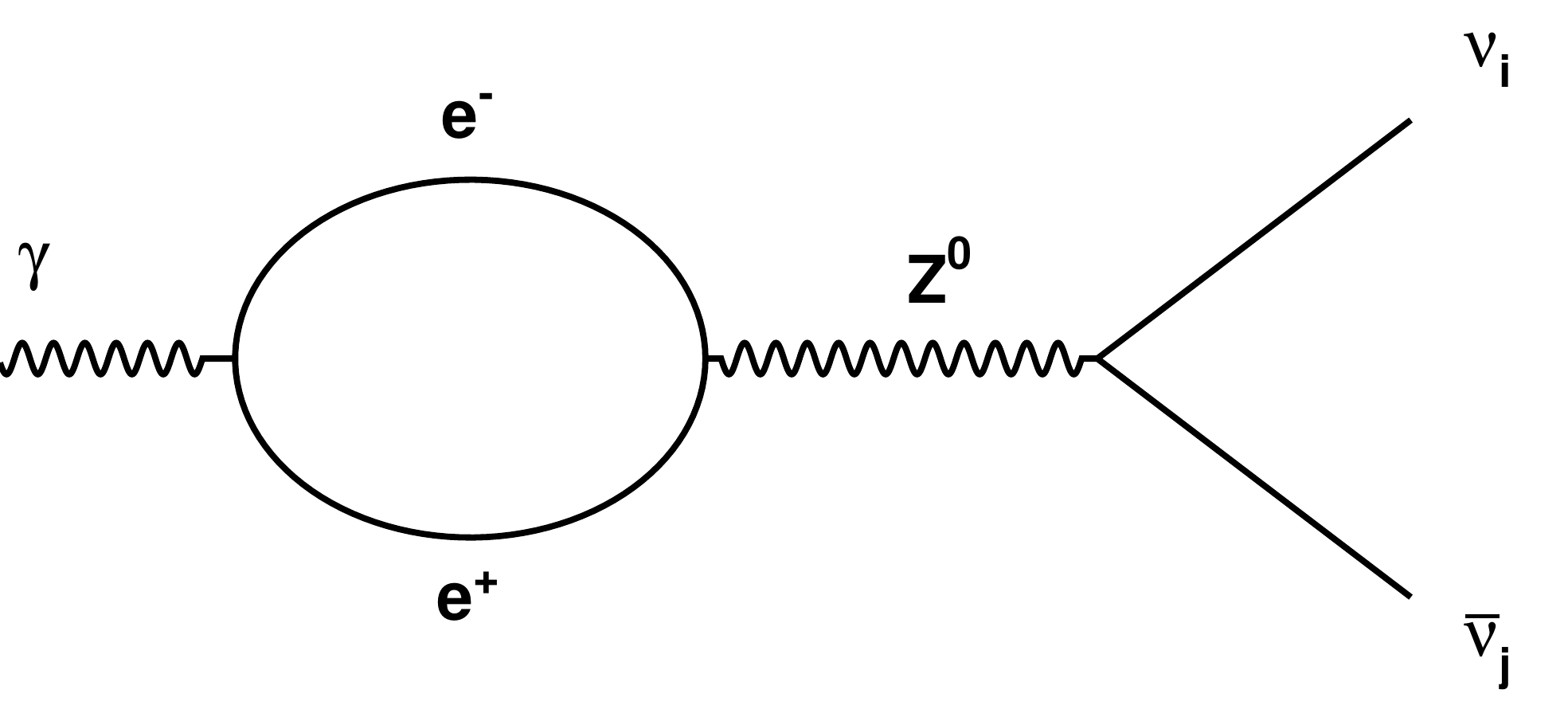}
    \caption{Diagrams that contribute to the plasmon decay: trough the charged and neutral currents.}
    \label{decay_diagrams_SM}
\end{figure}
Fig.~\ref{decay_diagrams_SM} displays the two SM diagrams contributing to plasmon decay. These diagrams can be expressed in a way that includes the previously computed self-energy,
\small{
\begin{align}
\mathcal{M} &= \frac{G_F}{\sqrt{8 \pi \alpha}} \Bigg[ \varepsilon_\mu (\omega_l,q) C_V^{\rm SM} \bigg(\Pi_{L} (\omega_l,q) \Big(1, \frac{\omega_l}{q} \hat{q} \Big)^\mu \Big(1, \frac{\omega_l}{q} \hat{q} \Big)^\nu \bigg) \nonumber\\
%&
&+ \varepsilon_\mu (\omega_t,q) g^{\mu i} \bigg( C_V^{\rm SM} \Pi_{T} (\omega_t,q) \Big(\delta^{ij} - \hat{q}^i \hat{q}^j \Big) \nonumber\\
&+ C_A   \Pi_{A} (\omega_t,q) (i \varepsilon^{ijm} \hat{q}^m) \bigg) g^{\nu j} \Bigg] \overline{u}(p_1) \gamma_\nu (1 - \gamma_5) v(p_2)
\end{align}}
$G_F $
is the Fermi constant, which appears  because $M_W^2(M_Z^2) \gg p^2$. 
The coefficient $C_V^{\rm SM}$ takes on a value of $2 \sin^2 \theta_W + 1/2$ for $\nu_e$, and $2 \sin^2 \theta_W - 1/2$ for other neutrino species. Meanwhile, $C_A$ is equal to $1/2$ for $\nu_e$ and $-1/2$ for all others. We are explicitly stating that $C_V$ belongs to the SM to distinguish it from the DS contribution. 
This is not the case for the axial component, as it remains unaffected by the inclusion of new physics. It should be noted that we assume the masses of the weak bosons to be much larger than the momenta involved in the process. Finally, if $\lambda=l,\, t$ represents the longitudinal  or transverse polarization, respectively, the above equation can be expressed using shorthand notation
\begin{equation}
\begin{split}
\mathcal{M} = \frac{G_F}{\sqrt{2}} \Big( \Gamma^{\mu \nu}_\lambda \varepsilon_\nu (\vec{q}, \lambda)  \Big) \overline{u}(p_1) \gamma_\nu (1 - \gamma_5) v(p_2)
\end{split}
\end{equation}
where a sum over polarisation is assumed. 

The decay width of the process given a specific polarization $\lambda$ is,
\begin{equation}
\begin{split}
\Gamma_\lambda (q) = \frac{1}{2 \omega_\lambda (q)} &\int \frac{d^3 p_1}{(2 \pi)^3} \frac{1}{2 p_1} \int \frac{d^3 p_2}{(2 \pi)^3} \frac{1}{2 p_2} \\&(2 \pi)^4 \delta^{(4)}(P_1 + P_2 - Q) |\mathcal{M}|^2.
\end{split}
\end{equation}
The integral can be performed easily since the final states only involve neutrinos, which are independent of the self-energy
\begin{equation}
\Gamma_\lambda (q) = -\frac{G_F^2}{12 \pi} \frac{\omega_\lambda (q)^2 - q^2}{\omega_\lambda (q)} \Big(\Gamma^{\alpha \mu}_\lambda \varepsilon_\mu (q, \lambda)\Big) \Big(\Gamma_{\alpha \rho}^\lambda \varepsilon^\rho (q, \lambda)\Big)^{*}
\end{equation}
Since the expression is evaluated at $q^0 = \omega_\lambda (q)$, it is possible to  use the dispersion relations to derive more explicit relations for each polarisation,
\begin{align}
\Gamma_l (q) &= (C_V^{\rm SM})^2 \frac{G_F^2}{48 \pi^2 \alpha} Z_l (q) \Big(\omega_l(q)^2 - q^2\Big)^2 \omega_l(q)\\
\Gamma_t (q)& = \frac{G_F^2}{48 \pi^2 \alpha} Z_t (q) \frac{\omega_t(q)^2 - q^2}{\omega_t(q)}\nonumber \\
&\times \bigg( (C_V^{\rm SM})^2 \Big(\omega_t(q)^2 - q^2\Big)^2 + C_A^2 \Pi_A (\omega_t(q),q)^2 \bigg)
\end{align}

\subsubsection{Emissivity of the plasma}
To calculate the Emissivity $\mathcal{Q}$ of the plasma, which represents the rate of energy loss per unit volume, we must integrate the decay rate over the phase space of the photon, with weighting by number density and energy, and sum over the polarization states of the photon as well as the different species of neutrinos. Upon substitution of the previously derived expressions, we obtain the following Emissivity for each type of polarization %
{\small
\begin{align}
\mathcal{Q}_T = & \sum_{\nu} (C_V^{\rm SM})^2 \frac{G_F^2}{48 \pi^4 \alpha}\!\! \int_0^\infty \!\!\!\!\!\!\!dq q^2 Z_t(q) \Big(\omega_t(q)^2 - q^2\Big)^3 \!\!\!n_B (\omega_t(q)) \\
\mathcal{Q}_A = & \sum_{\nu} C_A^2 \frac{G_F^2}{48 \pi^4 \alpha}\!\! \int_0^\infty \!\!\!\!\!\!\! dq q^2 Z_t(q)\Big(\omega_t(q)^2 - q^2\Big)\nonumber \\&~~~~~~~~~~~~~~~~~~~~~~~~~~~\times  \Pi_A (\omega_t(q),q)^2 n_B (\omega_t(q)) \\
\mathcal{Q}_L = &\sum_{\nu} (C_V^{\rm SM})^2 \frac{G_F^2}{96 \pi^4 \alpha} \!\!\int_0^\infty\!\!\!\!\!\!\! dq q^2 Z_l(q) \Big(\omega_l(q)^2 - q^2\Big)^2 \nonumber\\
&~~~~~~~~~~~~~~~~~~~~~~~~~~~\times\omega_l(q)^2n_B (\omega_l(q))
\end{align}}the sum of the squares of the vector and axial-vector couplings, represented by 
$C_V^{\rm SM}$ and $C_A$ respectively, overall neutrino species,  is given by $\sum_{\nu} 
(C_V^{\rm SM})^2 = 3/4 - 2 \sin^2 \theta_W + 12 \sin^4 \theta_W \approx 0.911$, while 
$\sum_{\nu} C_A^2 = 3/4$. The variable $ n_B$ represents the distribution function of the photons within the system.

To calculate the neutrino luminosity, the Emissivity must be integrated over the entire volume of the star. Assuming spherical symmetry, this can be expressed as
\begin{equation}
    \mathcal{L}_\Lambda = 4 \pi \int_{0}^{R_\star} \mathcal{Q}_\Lambda (r) r^2 dr\label{eq:luminosity}
\end{equation}
where $\Lambda$ stands for $L$, $T$ or $A$. $R_\star$ is the radius of the WD core, which is the volume considered for the plasmon decay. The quantities used to compute the emissivities depend simultaneously on the radius of the WD through its density or $p_F$, depending on the regime. For high chemical potential, the high degenerate approximation can be used. As we approach the surface of the core, the non-relativistic approximation is used instead. 

%%%%%%%%%%%%%%%%%%%%%%%%%%%%%
\subsection{Emission rate and the dark sector}
\begin{figure}
    \centering
    \includegraphics[width=0.35\textwidth]{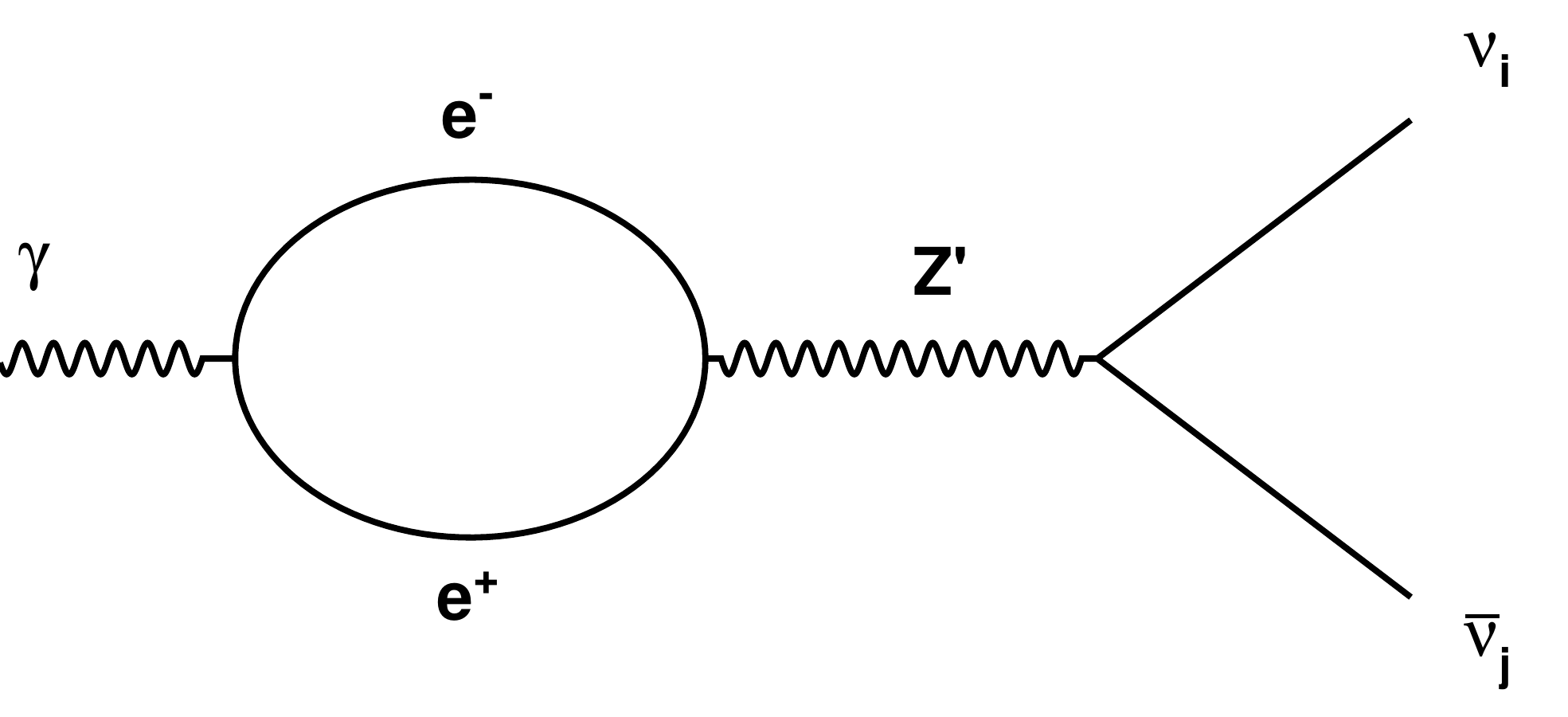}
    \caption{Diagram that contributes to the plasmon decay through the dark photon.}
    \label{decay_diagrams_dark}
\end{figure}

%%%%%%%%%%%%%%%%%%%%%%%%%%%%%
If dark photons are present and  interact with both an electromagnetic current and a current of dark neutrinos, it introduces the possibility of an additional diagram in the system that we sum to the SM  ones. The new interaction is analogous to the diagram involving the $Z$ particle, but instead involving the $Z'$ particle, as depicted in Fig.~\ref{decay_diagrams_dark}.  In this scenario, the dark final states would be mixed with the light states, which can be expressed in terms of the SM neutrinos. 

It is important to note that the photon/plasmon could not produce heavier mass states at the energies characteristic of a WD if we consider those heavy neutral leptons at the MeV scale. The plasma frequencies of a WD are well below those energies. 
In this case, we will not use the effective propagator of the dark photo,  $(Q^2 - M_{Z^\prime}^2)^{-1} \sim -M_{Z^\prime}^{-2}$, just to consider a general approach. The amplitude to neutrino mass states $i$ and $j$, coming from the diagram in Fig.~\ref{decay_diagrams_dark} is,
\begin{equation}
\begin{split}
\mathcal{M}^{ij}_{Z^{\prime}} &= -\varepsilon_\mu (Q) \int \frac{d^4k}{(2 \pi)^4} \mathrm{tr} \big[ \gamma^\mu S_\beta^F (K) \gamma^\nu S_\beta^F (K-Q) \big] \\&\times\epsilon e^2 \frac{1}{Q^2 - M^2_{Z^{\prime}}} \frac{g_\mathrm{D}}{2} U^{*}_{i \mathrm{D}} U_{j \mathrm{D}} [\overline{u}_j (p_1) \gamma_\nu (1 - \gamma_5) v_i(p_2)] \\
&= \frac{G_F}{2\sqrt{4 \pi \alpha}} \Big[C_{\nu}^\mathrm{D} U^{*}_{i \mathrm{D}} U_{j \mathrm{D}} \Pi^{\mu \nu} \Big] \overline{u}_j (p_1) \gamma_\nu (1 - \gamma_5) v_i(p_2)
\end{split}\label{eq:DS_amp}
\end{equation}
where $U$ represents the mixing matrix for neutrino states, with $U_{i\mathrm{D}}$ denoting the mixing between a dark state $D$ and a light mass state $i$. Here  $S_\beta^F (P)$ is the thermal fermion propagator for the electron/positron at temperature $T \equiv 1 / \beta$. The constant $C_{\nu}^\mathrm{D}$ is defined as
\begin{equation}
C_{\nu}^\mathrm{D} = \frac{\sqrt{2 \pi \alpha}}{G_F} \frac{\epsilon g_\mathrm{D}}{M^2_{Z^{\prime}} - Q^2}.
\end{equation}
When computing the neutrino emission rates, it is convenient to express each  contribution in terms of neutrino mass states, $i$ and $j$, rather than flavour states in order to observe the interference with the DS. 
 These mass states may not necessarily be the same, and hence, the summation must be performed over all possible combinations,
\begin{align}
C_{V,ij}^\mathrm{SM+D} = &\sum_\alpha C_{V,\alpha}^{\mathrm{SM}} U_{\alpha i}^* U_{\alpha j} + C_{\nu}^\mathrm{D}  U_{\mathrm{D} i}^* U_{\mathrm{D} j} \nonumber\\
\big(C_{V,ij}^\mathrm{SM+D}\big)^2& = \big|\sum_\alpha C_{V,\alpha}^{\rm SM} U_{\alpha i}^* U_{\alpha j}\big|^2+ |C_V^{\rm D} \,U_{\mathrm{D} i}^* U_{\mathrm{D} j}|^2\nonumber\\&
+ \frac{\sqrt{8 \pi \alpha}}{G_F} \frac{\epsilon g_\mathrm{D} U_{\mathrm{D} i}^* U_{\mathrm{D} j}}{M^2_{Z^{\prime}} - Q^2} \Re \Bigg[ \sum_\alpha C_{V,\alpha}^{\rm SM} U_{\alpha i}^* U_{\alpha j} \Bigg] 
\end{align}
where $C_{V,\alpha}^{\rm SM}$ is  for a particular flavor $\alpha$. Performing the sums over all possible final mass eigenstates, we compute the decay width,
\small{\begin{align}
\big(C_{V}^\mathrm{SM+D}\big)^2 &= \sum_{\alpha} \big(C_V^{\rm SM}\big)^2+ \frac{18 \pi \alpha}{G_F^2} \frac{\epsilon^2 g_\mathrm{D}^2 \big|U_{\mathrm{D}}\big|^4}{(M^2_{Z^{\prime}} - Q^2)^2}\nonumber \\
&+ \frac{\sqrt{8 \pi \alpha}}{G_F} \frac{\epsilon g_\mathrm{D} \big|U_{\mathrm{D}}\big|^2 }{M^2_{Z^{\prime}} - Q^2} \Re \Bigg[ \sum_{\alpha,i,j} C_{V,\alpha}^{\rm SM} U_{\alpha i}^* U_{\alpha j} \Bigg].
\label{eq:SM_DS_CV}
\end{align}}Here, $(C_{V}^\mathrm{SM+D})^2\equiv\sum_{\small ij}(C_{V,\,ij}^\mathrm{SM+D})^2$ has contributions from three terms. The first term represents the SM contribution, while the second term is solely from the DS. The third term represents the interference between the SM and DS amplitudes. 

To simplify the computation, the assumption has been made that $U_{\mathrm{D} i}$ is equal for every $i$, denoted as $U_{\mathrm{D}}$. This implies that the mixing between each light mass state and the dark states is real and equal without any loss of generality. With the inclusion of the dark photon in the photon/plasmon decay, the transverse and longitudinal emissivities are also altered as a consequence,
\begin{align}
\mathcal{Q}_T = 2 \frac{G_F^2}{96 \pi^4 \alpha} &\int_0^\infty dq q^2 Z_t(q) \sum_{\alpha\beta} \big( C_V^{\rm SM+D} (\omega_t(q),q)\big)^2 \nonumber \\&\times\Big(\omega_t(q)^2 - q^2\Big)^3 n_B (\omega_t(q)),\label{eq:Qt_dark} \\
\mathcal{Q}_L = \frac{G_F^2}{96 \pi^4 \alpha} &\int_0^\infty dq q^2 Z_l(q) \sum_{\alpha\beta} \big( C_V^{\rm SM+D} (\omega_l(q),q)\big)^2 \nonumber\\&\times \omega_l(q)^2 \Big(\omega_l(q)^2 - q^2\Big)^2 n_B (\omega_l(q)).\label{eq:Ql_dark} 
\end{align}
It is evident from Eq.~\eqref{eq:SM_DS_CV},  that the dependence on $G_F^2/\alpha$ vanishes for the purely DS contribution. Notice that 
$\mathcal{Q}_A$ remains unchanged because there is no axial term next to the thermal loop in the dark photon diagram (see Fig. \ref{decay_diagrams_dark}). 

Finally, using Eq.~\eqref{eq:luminosity}, \eqref{eq:Qt_dark} and \eqref{eq:Ql_dark}, we can compute the WD luminosity due to photon/plasmon decay into neutrinos.

%%%%%%%%%%%%%%%%%%%%%%%%%%%%%
\section{Results and discussion}
\label{sec:results}
%%%%%%%%%%%%%%%%%%%%%%%%%%%%%

To obtain the total luminosity of a WD due to photon/plasmon decay is necessary to obtain the WD radial profiles. This is achieved through the use of the Tolman-Oppenheimer-Volkoff (TOV) equations~\cite{mathew2014general}, which describe the hydrostatic equilibrium of a spherically symmetric, non-rotating star coupled to  the Salpeter EOS~\cite{salpeter61}. It is important to note that the following analysis is performed on a hypothetical young WD with  1 solar mass.  We fix its temperature to be of the order of $10^8$ K, and using the evolutionary sequences given in Ref.~\cite{2020ApJ...901...93B}, this temperature corresponds to a WD of $\sim\mathcal{O}(10^6)$ s old.

It is necessary to establish some limits to determine the parameter space of the three-portal model that will be explored. One such limit concerns the quantity $\epsilon g_\mathrm{D} |U_\mathrm{D}|^2$. Here, $|U_\mathrm{D}|$ refers to $|U_\mathrm{D i}|$, where $i$ denotes a massive light neutrino state. Current limits on this element depend on the mass scale of the heavy states, and we want to avoid the heavy states being so light that they can be directly produced by the plasmon.\footnote{ Although this effect might increase the energy lost by plasmon decay.} Based on Ref.~\cite{Flieger:2019eor}, we can assume that $|U_\mathrm{D}| \lesssim 10^{-1}$. This would imply high masses for the heavy neutrino states. We are not particularly interested in the specific realization of this as long as the values are not fully excluded. The parameter $g_\mathrm{D}$ is not a coupling of the dark photon to SM fermionic currents, so limits such as those found in~\cite{Billard:2018jnl} are not relevant for our model. There are no direct constraints on $g_\mathrm{D}$ neither if we do not consider the dark photon or the HNLs to be dark matter, we still need to keep the theory perturbative on $\alpha_\mathrm{D} \equiv g_\mathrm{D}^2 / 4\pi$. Therefore, we can assume that $g_\mathrm{D} \sim \mathcal{O}(1)$. The excluded regions for $\epsilon$ depend heavily on the mass of the dark photon and the number of extra neutrino states. We can assume $\epsilon \lesssim 10^{-2}$ based on~\cite{Mongillo:2023hbs,Abdullahi:2023tyk} which is safe, especially for heavy neutrino mass states of the order of $1-10$ GeV~\cite{Marocco:2020dqu}. Hence, we can consider $\epsilon g_\mathrm{D} |U_\mathrm{D}|^2 \lesssim 10^{-4}$.

On the other hand, for the mass of the dark photon, the limits depend on several conditions: whether the dark photon or the heavy neutral leptons constitute dark matter, the value of $\epsilon$ or $g_\mathrm{D}$ and the mixing matrix of the neutrino sector, $U$. Here, we will also consider the same parameter space of~\cite{Mongillo:2023hbs,Abdullahi:2023tyk}, such that $10\, \mathrm{ MeV} \leq M_{Z\prime} \leq 10\, \mathrm{ GeV}$. The Lagrangian in Eq.~\eqref{darkphotoninteraction} is valid as long as $\big( M_{Z\prime} / M_{Z} \big)^2$ is negligible. Since for $M_{Z\prime} = 10\, \mathrm{ GeV}$, this is of the order of $10^{-2}$, and its contribution is still negligible.

\subsection{Luminosity}
\begin{figure}
    \centering
    \includegraphics[width=0.48\textwidth]{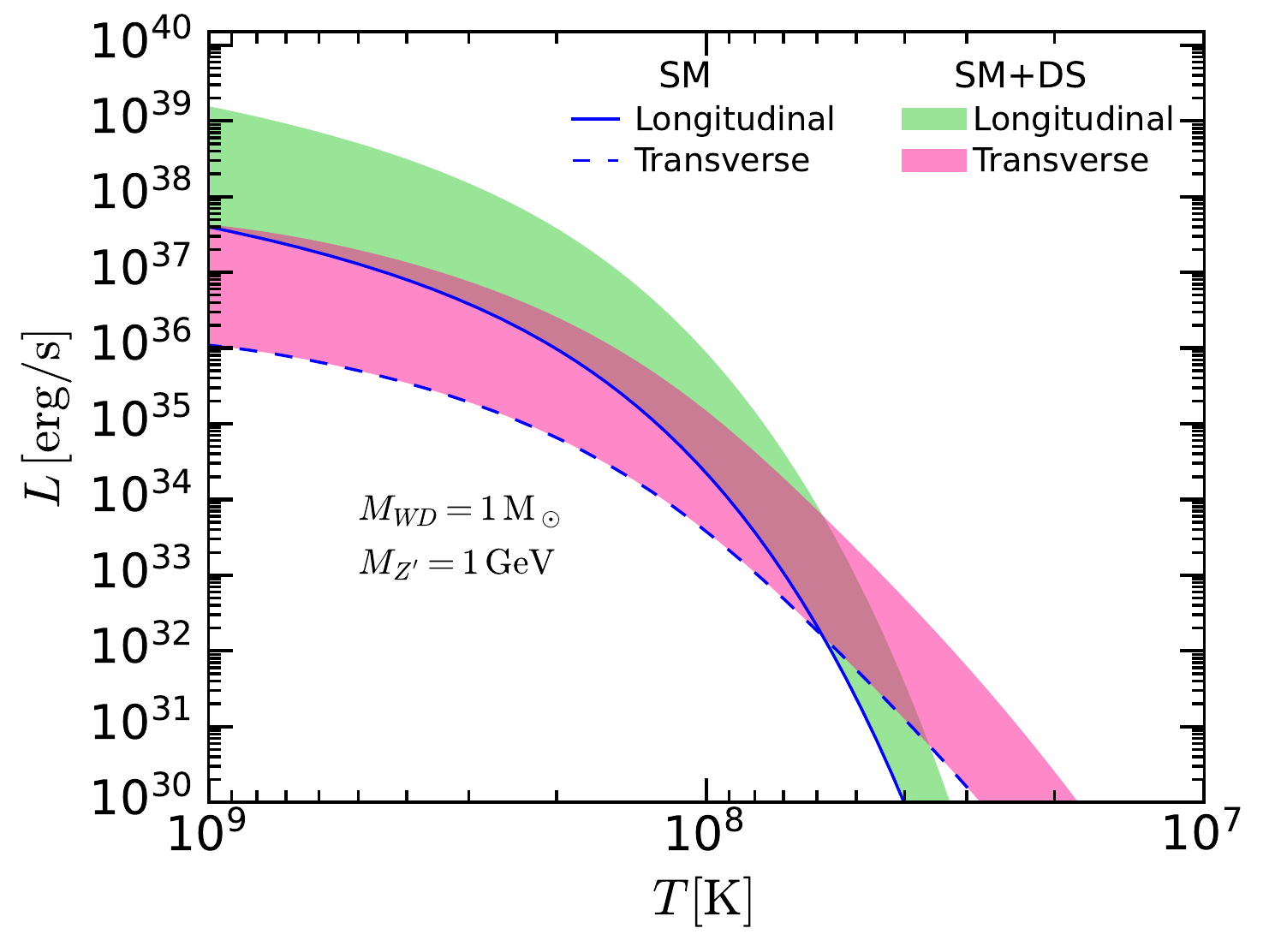}
    \caption{Luminosity of a $1 M_\odot$ young WD with respect to its temperature for different BSM scenarios. The solid (dashed)  line represents  the longitudinal (transverse) components. The green (pink) bandwidth of the DS longitudinal (transverse) contribution considers the values: $\epsilon g_D |U_D|^2=[10^{-8}-10^{-4}]$. The axial contribution is not just shown since no contribution comes from the new physics.}
    \label{L_vs_T}
\end{figure}

Fig.~\ref{L_vs_T} illustrates  the total luminosity of the WD attributed solely  to plasmon decay.  The plot  depicts different scenarios. The blue solid (dashed) lines correspond to the longitudinal (transverse) contributions of the SM case. In contrast, the green and pink regions show the longitudinal and transverse contributions of the DS scenario, which includes a dark photon with a mass of $M_{Z^\prime}=1$ GeV. Both bands demonstrate the luminosity for two different coupling strengths: $\epsilon g_D |U_D|^2=10^{-8}$  and $\epsilon g_D |U_D|^2=10^{-4}$. Note that when the coupling is too small, the curves correspond to the SM scenario. The maximum luminosity from new interactions in the neutrino emission rate is about one order of magnitude greater than the SM. This  suggests that dark photons, like the one proposed in the DS scenario, may contribute to the evolution and behaviour of WDs in ways not accounted for by the SM. As the WD cools down,  the contributions from plasmons decrease until they essentially disappear. Before reaching a temperature of $10^8$ K, the main source of energy loss is from the longitudinal contribution (solid line). At lower temperatures, the transverse contribution (dashed line) dominates. Here, we do not  show the axial contribution in our analysis since it is suppressed several orders of magnitude  compared to the longitudinal and transverse contributions. 

\begin{figure}
    \centering
    \includegraphics[width=0.48\textwidth]{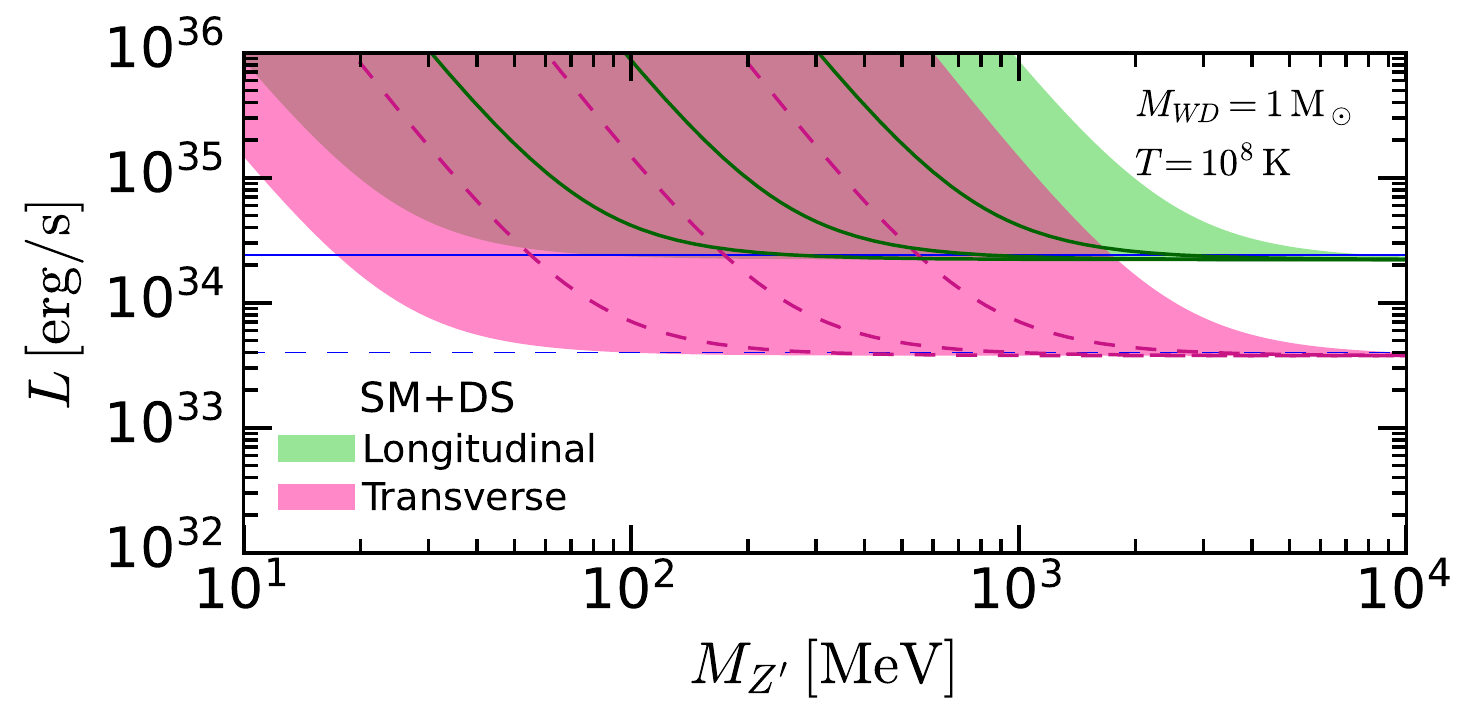}
    \caption{Luminosity of a $1 M_\odot$ young WD with a temperature of $T=10^8$ K, as a function of the dark photon mass for different BSM scenarios. The solid (dashed) line represents the longitudinal (transverse) components. The axial contribution is not shown since there is no contribution from new physics. The bandwidth of the DS contribution considers the values: $\epsilon g_D |U_D|^2=[10^{-8}-10^{-4}]$.}
    \label{L_vs_mZp}
\end{figure}
Fig.~\ref{L_vs_mZp} shows the luminosity of the WD, at a temperature of $T = 10^{8}$ K, as a function of the dark photon mass $M_{Z^\prime}$. Again, the solid (dashed) blue line represents the SM  longitudinal (transverse) contribution. The green and pink areas correspond to the emission rates, including the dark photon, and the bands represent various fixed values of $\epsilon g_D |U_D|^2=[10^{-8}-10^{-4}]$.  As expected, for each set of couplings, there is an upper bound on the dark photon mass above, which the new physics does not visibly contribute, and the luminosity curve is just a horizontal line approaching the SM. As the dark photon mass decreases, the contribution to the luminosity grows exponentially, with the dominant contribution coming from the longitudinal photon states (solid lines). For instance, when the product of couplings is $\epsilon g_D |U_D|^2 = 10^{-8}$, and the dark photon has a mass of $\sim10$ MeV, the luminosity of the WD can be up to $\sim 2$ orders of magnitude higher than the SM case. This effect becomes insignificant for $M_{Z^\prime}\sim 100$ MeV and above. On the other hand, in the case of $\epsilon g_D |U_D|^2 = 10^{-4}$, the effect of the new interaction in the WD luminosity becomes irrelevant around $10$ GeV.  
Notice that as the product $\epsilon g_D |U_D|^2$ increases, the impact of the new physics in the luminosity is visible for heavier dark photon states. This can be seen from the green solid and pink dashed lines showing, from left to right, $\epsilon g_D |U_D|^2 = 10^{-7}, 10^{-6}$ and $10^{-5}$.

\subsection{Results}
\begin{figure}
    \centering
    \includegraphics[width=0.48\textwidth]{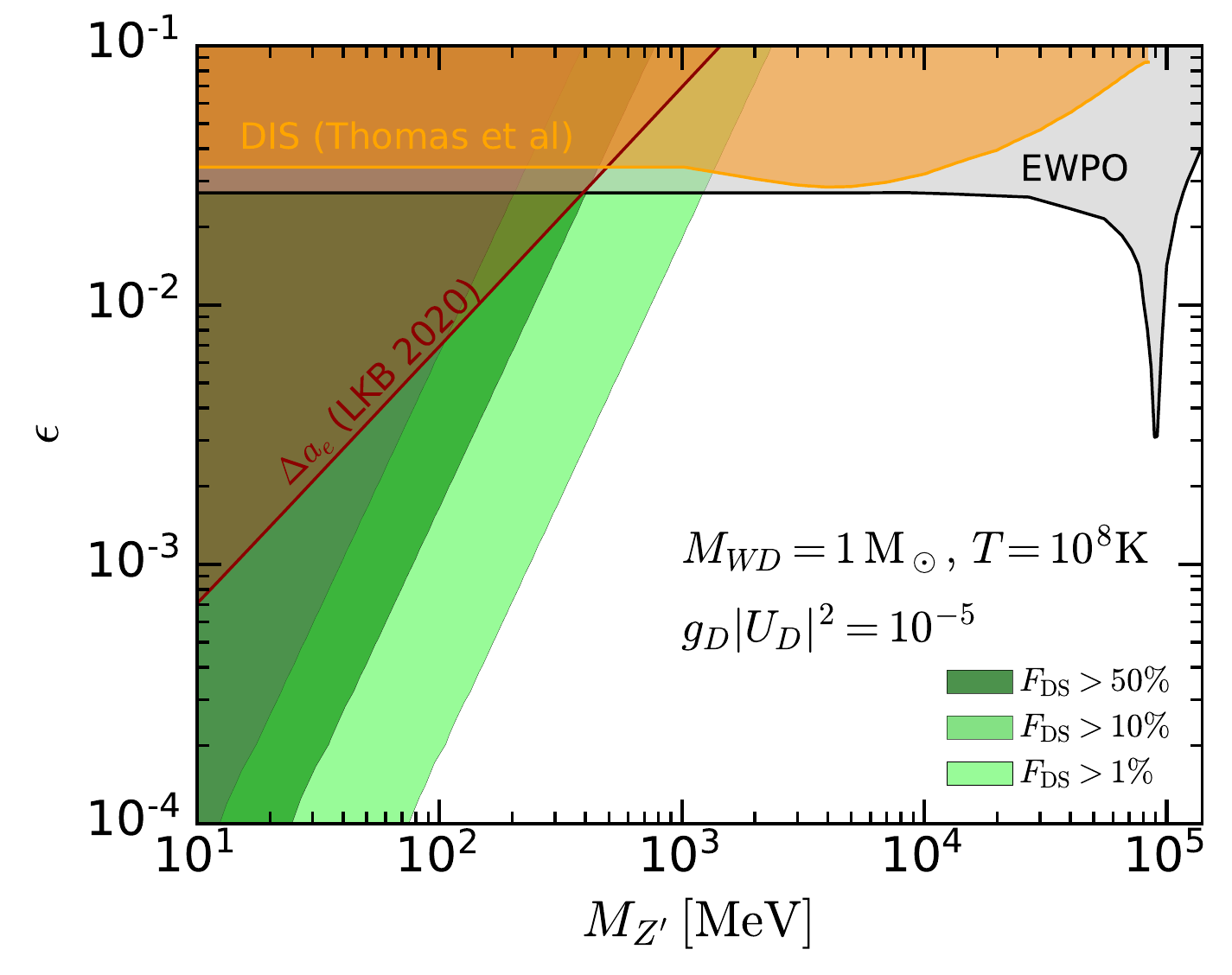}
    \caption{Limits on  $\epsilon$ for dark photons decaying into neutrinos, obtained with a young WD assuming $T=10^8$ K and a dark sector parameter of $g_D|U_D|^2=10^{-5}$. The luminosity due to dark photons constitutes a maximum of $F_{DS}=1\%$ (lightest green shaded area), $10\%$, and $50\%$ (darkest green shaded area) of the total SM luminosity. For comparison, we also show the bounds from DIS~\cite{Thomas:2021lub}, electron ($g-2$) (LKB)~\cite{2018Sci...360..191P} and \cite{Curtin:2014cca} proportioned in Ref.~\cite{Abdullahi:2023tyk}. }
    \label{couplings_vs_mZp}
\end{figure}Finally, we present a comprehensive investigation of the parameter space associated with the dark photon in the context of the cooling of WDs.  Here, we express the contribution of dark photons as a fraction of the total SM luminosity,
\begin{equation}
    F_{\mathrm{DS}} = \frac{\mathcal{L}_\mathrm{DS + SM} - \mathcal{L}_\mathrm{SM}}{\mathcal{L}_\mathrm{SM}} \times 100 \%.
\end{equation}
Therefore,  we can perform estimations and projections of the allowed parameter space regarding the cooling of WD mediated by dark photons.

By imposing the condition that the luminosity generated through the presence of dark photons constitutes a maximum of $1\%$, $10\%$, and $50\%$ of the total SM luminosity, we can determine the excluded parameter space.  In  Fig.~\ref{couplings_vs_mZp}, we set reasonable dark sector parameters  $g_D|U_D|^2=10^{-5}$, and show the estimated bounds on $\epsilon$. The bounds are depicted as shaded regions, ranging from the lightest to the darkest shade of green, corresponding to $1\%$, $10\%$, and $50\%$ maximum values of $F_{DS}$, respectively. For comparison, we also show the bounds given by Deep-inelastic scattering (DIS)~\cite{Thomas:2021lub}, electron ($g-2$) (LKB)~\cite{2018Sci...360..191P} and electroweak precision observables (EWPO)~\cite{Curtin:2014cca}, revised and presented in \cite{Abdullahi:2023tyk} as model-independent limits. 

We notice that in all cases, the estimated bounds significantly surpass the existing constraints. As the contribution of dark photons decreases, the bound becomes progressively more stringent. This behaviour can be attributed to the fact that, in order to achieve a smaller DS contribution, the coupling parameter $\epsilon$ must decrease accordingly. The obtained bounds are up to  one and two orders of magnitude, larger than those given by precision measurements of the electron anomalous magnetic moment when considering  a maximum  fraction of $F_{DS}=50\%$ and $10\%$, respectively. However, the most compelling result emerges when we restrict the luminosity to only $1\%$ of the SM luminosity. In this scenario, the range where the bound exhibits its greatest strength extends from 10 MeV to 1 GeV. It is important to note that we are unable to  extend our bounds beyond 10 GeV since the  validity of the dark photon interactions taken into account relies on the assumption that $\big( M_{Z\prime} / M_{Z} \big)^2<<1$. Hence, for $M_{Z^\prime}>10^4$ MeV, the strongest bounds are still given by EWPO and DIS. 

There are certainly additional constraints on this parameter space~\cite{Essig:2010gu,Ilten:2018crw}. However, in order to translate these bounds, we would need more specific considerations on the model, especially those that affect the visible, semi-visible and invisible decays of the dark photon. Since our computation is independent of those regards, we are not showing them here. Finally, it is worth noting that electron-neutrino scattering can be mediated through the exchange of a dark photon. In the context of B-L models, the constraints imposed on this process are more stringent compared to those derived from white dwarf cooling. This is primarily due to the fact that in B-L models, the vector boson couples with the same strength to the SM and the DS. On the other hand, in  the case of the three-portal model, the couplings are independent, resulting in weaker constraints from the electron-neutrino scattering process.
\section{Conclusions}
\label{sec:conclusions}

In the present paper, we have computed the total luminosity of a white dwarf due to photon/plasmon decay and present various scenarios involving a dark photon with different masses and couplings.  We found that for a $M_{Z^\prime}= 1$ GeV, the maximum luminosity from new interactions, corresponding to $ \epsilon g_D |U_D|^2=10^{-4}$, exceeds that of the SM by approximately one order of magnitude. Additionally, we have examined  the upper bound on the dark photon mass for  each set of couplings.  Above this mass,  the new physics has a minimal visible contribution, and the luminosity from new interactions aligns with the SM one. We found that in the  case of  $ \epsilon g_D |U_D|^2=10^{-4}$, the dark photon mass at which the luminosity becomes only SM luminosity is around 10 GeV. 

Finally, by  imposing the condition that the luminosity attributed to dark photons should not exceed $1\%$, $10\%$, and $50\%$ of the total Standard Model luminosity, we  estimated bounds on the coupling parameter $\epsilon$ for reasonable dark sector parameters $g_D|U_D|^2=10^{-5}$. Remarkably, our estimated bounds consistently surpass the existing constraints, becoming increasingly stringent as the contribution of dark photons decreases. This is due to the need for smaller coupling values of $\epsilon$ to achieve a reduced dark sector contribution. Notably, when restricting the luminosity to only $1\%$ of the SM luminosity, our bounds exhibit their strongest range from 10 MeV to 1 GeV. Therefore, the cooling behaviour of WDs presents a promising avenue to probe the existence and properties of dark photons. 

\acknowledgments
We thank Matheus Hostert for the helpful discussion on the dark sector model and current bounds. We also thank Daniele Massaro for providing the revised bounds from EWPO, LKB and DIS.
MRQ would like to thank Koichi Hamaguchi for  the useful discussion and feedback. 
The research of JHZ has received support from the European Union’s Horizon 2020 research and innovation programme under the Marie Sk\l{}odowska-Curie grant  agreement No 860881-HIDDeN. 
MRQ is supported  by the JSPS KAKENHI Grant Number 20H01897.

\bibliographystyle{apsrev4-1}
\bibliography{lib}{}

%%%%%%%%%%%%%%%%%%%%%%%%%%%%%%%%%%%%%%%%
% supplemental materials
\pagebreak
\appendix
\onecolumngrid
%%%%%%%%%%%%%%%%%%%%%%%%%%%%%%%%%%%%%%%%

\end{document}